# EEG Signal Denoising Using pix2pix GAN: Enhancing Neurological Data Analysis

Haoyi Wang, Xufang Chen, Yue Yang, Kewei Zhou, Meining Lv, Dongrui Wang, Wenjie Zhang

## Abstract

Electroencephalography (EEG) is essential in neuroscience and clinical practice, yet it suffers from physiological artifacts, particularly electromyography (EMG), which distort signals. We propose a deep learning model using pix2pixGAN to remove such noise and generate reliable EEG signals. Leveraging the EEGdenoiseNet dataset, we created synthetic datasets with controlled EMG noise levels for model training and testing across a signal-to-noise ratio (SNR) from -7 to 2. Our evaluation metrics included RRMSE and Pearson's CC, assessing both time and frequency domains, and compared our model with others. The pix2pixGAN model excelled, especially under high noise conditions, showing significant improvements in lower RRMSE and higher CC values. This demonstrates the model's superior accuracy and stability in purifying EEG signals, offering a robust solution for EEG analysis challenges and advancing clinical and neuroscience applications.

Keywords: EEG, EMG, pix2pixGAN, denoising

## 1. INTRODUCTION

Electroencephalography (EEG) is an important non-invasive neuroimaging technique that records the electrical activity generated by neuronal discharges in the brain through electrodes placed on the scalp. It plays a crucial role in various neurological research and therapeutic applications such as the detection of seizures, diagnosis of Alzheimer's disease, sleep stage analysis, brain-computer interfaces (BCIs), and biometric systems[1][2][3][5][6]. However, EEG recordings are highly susceptible to contamination by a variety of physiological[7][8][9], and non-physiological artifacts that may significantly affect the accuracy of EEG data analysis and even lead to false diagnoses in severe cases. Among these artifacts, electromyographic (EMG)[10] noise originating from muscle activity (e.g., muscle contractions and movements near the scalp), is particularly difficult to remove. EMG artifacts, which typically occur in temporal and occipital lobe regions, have broad spectral content and overlap with the frequency bands of interest in EEG analysis, and may mask or distort the underlying neural signals, leading to an incorrect diagnosis. EMG artifacts in EEG signals are The prevalence of EMG artifacts in EEG signals creates an urgent need for the development of accurate and effective methods for estimating and eliminating these noisy signals to obtain high-quality EEG recordings suitable for reliable analysis and diagnosis[11][12][13].

Traditional denoising techniques have been widely explored for removing artifacts from EEG data. Methods such as regression-based approaches[15][16], adaptive filtering[17], and blind source separation (BSS) techniques[18][19] such as independent component analysis (ICA) and typical correlation analysis (CCA) have been used to attenuate EMG artifacts. However, these methods usually rely on certain assumptions, such as the availability of a reference signal or specific statistical properties of the artifacts, which may not always hold in practical situations. In addition, some techniques require a large number of electrodes or involve complex parameter tuning, limiting their applicability.

In recent years, deep learning (DL) has received increasing attention for its ability to learn complex patterns from large amounts of data without the need for manual features or a priori assumptions[20][21][22][23][24]. Deep neural networks have been successfully applied to various EEG signal analysis tasks, especially in denoising[25][26]. DL-based methods can learn the fundamental properties of neural oscillations in EEG signals and effectively distinguish them from artifacts such as EMG noise. Despite the progress made by DL methods in EEG denoising, several challenges remain. Previous studies have highlighted issues such as the lack of a robust framework, the limited generalization ability of the model, the use of selective portions of the dataset for testing, and poor models.

In the paper, we present a model for a reliable pix2pixGAN designed specifically for removing EMG artifacts from contaminated EEG signals. The model utilizes the generative power of the GAN to efficiently separate EMG noise from EEG signals, producing high-quality clean EEG data. We use a benchmark dataset, EEGdenoiseNet[14], to examine the performance of the model and ensure that our results are comparable and reproducible. We compare the denoising performance of our model with other models at different SNRs by several performance metrics, which indicate that our proposed GAN model is capable of high-quality denoising, especially in the most challenging case of denoising low SNR, which demonstrates far better denoising performance than other models. The superior performance of our model demonstrates its robustness and effectiveness in enhancing the quality of EEG signals, demonstrating its potential for practical deployment in clinical and research settings.

## 2. RELATED WORK

Generative Adversarial Networks (GANs) have made significant progress in the field of image generation and synthesis, initially demonstrating their powerful ability to generate realistic images. In recent years, more and more studies have applied GANs to the generation, interpolation, and enhancement of time series. In this study, we specifically use the pix2pixGAN model for denoising EEG signals. By converting clean and contaminated EEG signals into images, we utilize the advantages of GAN in image generation and improve signal quality.

Previous studies have explored the application of GAN in EEG signaling, with contributions from Luo and Lu[27], Sunhee Hwang et al[28], Ghaith Bouallegue et al[29], Fahimi et al[31], Prajwal Singh et al[30], Jin Yin et al[32], and Carlos de la Torre-Ortiz[33]. However, specific applications for denoising EEG signals are still limited. Notably, Gandhi et al.[34] developed an asymmetric GAN aimed at denoising EEG time-series data; their method uses unpaired training datasets and does not require explicit information about the noise source to achieve denoising. While these methods have shown some effectiveness in reducing noise in EEG signals, they typically do not focus on the removal of specific artifacts and lack a reliable quantitative assessment of signal-to-noise ratio (SNR) improvement. Our study fills this gap by demonstrating the pix2pixGAN model as a powerful artifact removal tool. By converting EEG signals to images, we utilize the advantages of GAN in image-to-image conversion to generate clearer and more reliable outputs. Through comprehensive benchmarking experiments, we provide qualitative and quantitative evidence of the model's performance, demonstrating its competitiveness with existing state-of-the-art methods in EEG artifact denoising.

In addition to GAN, other deep learning techniques such as convolutional neural networks (CNNs)[36][37][38] have been applied to EEG denoising with varying degrees of success. While these methods have proven effective in some cases, our results show that the pix2pixGAN model, which focuses on image transformation, performs particularly well in removing electromyographic (EMG) noise, thus providing a unique advantage over traditional methods.

In summary, the development of EEG denoising methods has witnessed the introduction of deep learning techniques, especially GAN. On this basis, our study improves noise removal and enhances the overall quality of EEG signal processing by using the pix2pixGAN model. By advancing recent advances in EEG artifact removal with this innovative image-based approach, we hope to provide valuable insights for more accurate interpretation of EEG data in clinical and research applications.

# 3. MATERIALS AND METHODS

## 3.1 Computational Tools and Software

In this study, we used MATLAB (MathWorks, Inc., Natick, MA, USA) to preprocess the public dataset. The pix2pixGAN model was run using PyCharm (JetBrains s.r.o., Prague, Czech Republic). Validation results were plotted using OriginPro (OriginLab Corporation, Northampton, MA, USA). The article was slightly polished using GPT-o1-mini (OpenAI, San Francisco, CA, USA).

## 3.2 pix2pixGAN

We use the pix2pixGAN model presented at CVPR 2017 by Phillip Isola et al.[4] The model utilizes a conditional generative adversarial network (cGAN), where the generator G uses the U-Net structure and the discriminator D uses a Markov discriminator (PatchGAN). The conditional GAN learns a mapping from an observation image x and a random noise vector z to a target image y, G:{x,z}→y. The generator G is trained to produce "true" images that are indistinguishable from the discriminator D, which is trained to recognize the "fake" images produced by the generator as accurately as possible. fake" images generated by the generator. The model architecture is shown in Fig. 4.

The objective function of cGAN can be expressed as:

$$\min_{G} \max_{D} L_{cGAN}(G, D) = E_{x,y}\left[\log D(x, y)\right] + E_{x,z}\left[\log\left(1 - D(x, G(x, z))\right)\right] \quad (1)$$

Where $L_{cGAN}(G, D)$ denotes the loss function of cGAN, and $D(x, G(x, z))$ denotes the probability that the discriminator D will judge the image $G(x, z)$ generated by the generator G as real. The discriminator D wants to maximize the accuracy of the determination of the true image while minimizing the probability of misdetermination of the generated image, maximizing $\log(1 - D(x, G(x, z)))$ by making $D(x, G(x, z))$ as small as possible. For discriminator D, the goal is to maximize this loss function. Correspondingly, generator G aims to generate images that can deceive the discriminator so that $D(x, G(x, z))$ is as large as possible, thus minimizing $\log(1 - D(x, G(x, z)))$, which means that for generator G, the goal is to minimize the loss function.

However, the goal of the generator G is not only to generate images that can deceive the discriminator D but also to generate images that go as close as possible to the real image y of the target domain.

$$L_{L1}(G) = E_{x,y,z}\left[\|y - G(x, z)\|_1\right] \quad (2)$$

Where $\|\ \|_1$ denotes the L1 paradigm. Therefore, the final objective function of the model is:

$$G^* = \arg\min_{G}\max_{D} L_{cGAN}(G, D) + \lambda L_{L1}(G) \quad (3)$$

The pix2pixGAN network we use is capable of mapping noisy time-series EEG signals into clean time-series EEG signals. In this study, we first convert the EEG time series containing electromyographic (EMG) noise and the corresponding clean EEG time series into images. Then, the noise-containing images are used as inputs to the generator to produce predicted images, and subsequently, these predicted images are fed into the discriminator along with the clean images to be discriminated by the discriminator. Through this training process, the ability of the generator to generate clean and denoised images is continuously enhanced. Finally, we convert the generated images back to EEG signals in the form of time series for further testing and analysis. The training process is shown in Fig. 1.

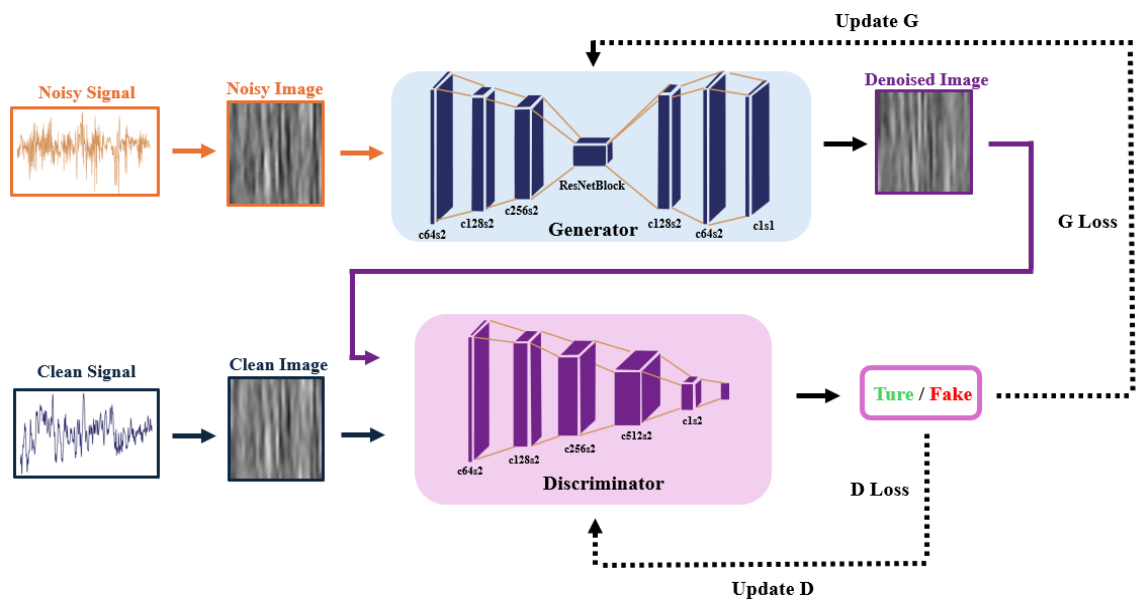

Figure 1 Overall architecture of the denoising model

## 3.3 Dataset Description

In this experiment, we used the publicly available EEGdenoiseNet dataset to validate our deep learning denoising technique. This dataset contains pre-processed and well-structured physiological signal fragments. Specifically, the authors of EEGdenoiseNet bandpass-filtered the motor imagery brain-computer interface data from 1 to 80 Hz, removed the power line noise by a trap filter, and then re-sampled the signal to 512 Hz to obtain 2-second-long signal segments containing 1024 data points each, yielding 4,514 signal segments. Each segment contains 512 data points, resulting in a total of 4,514 clean EEG clips. For the electromyography (EMG) dataset, bandpass filtering was applied from 1 to 120 Hz, followed by trap filtering at the power frequency, and the signal was resampled to 512 Hz. A total of 5,598 myogenic artifact segments were processed, each containing 1,024 data points. We synthesized these EMG noisy fragments with clean EEG clips to generate noise-contaminated EEG clips. These synthesized noise-containing EEG signals were then used to train and test our pix2pixGAN model as well as other deep learning models to perform a comparison of their denoising performance.

As shown in Fig. 2, we present a clean EEG clip, a clean EMG clip, and noise-contaminated EEG clips synthesized under different SNR conditions.

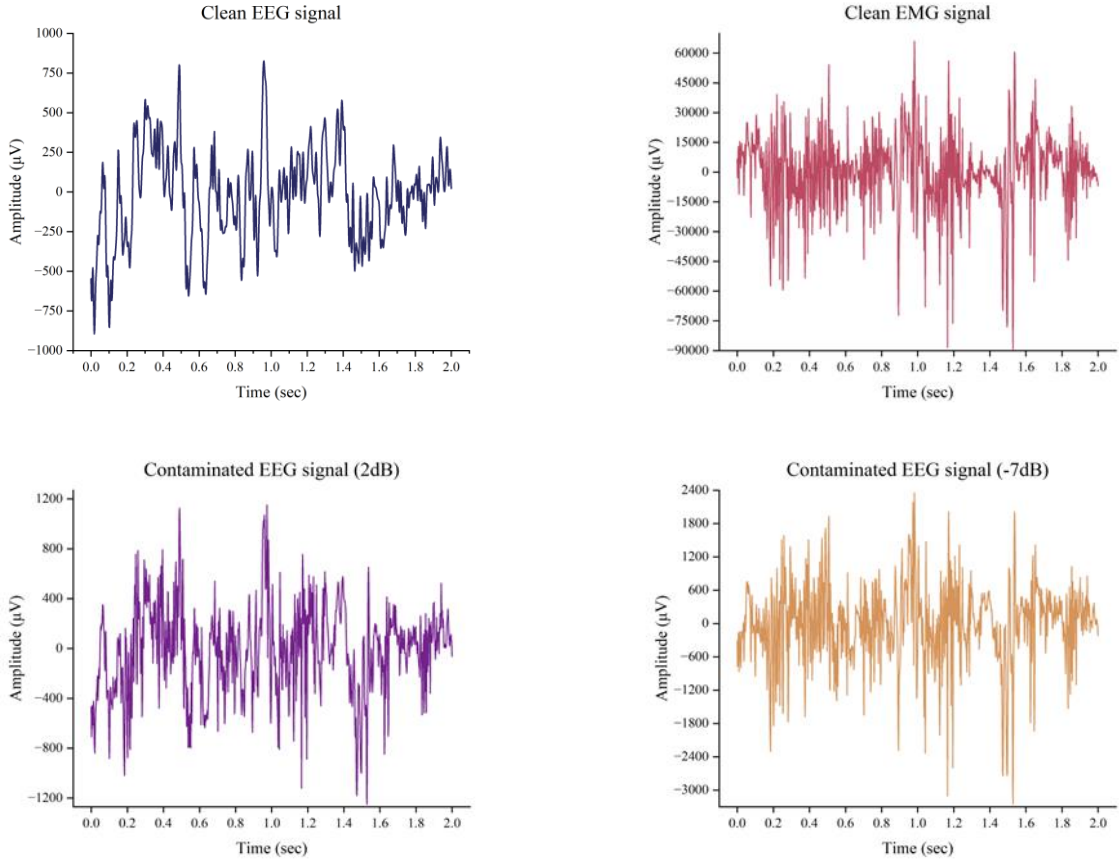

Figure 2 Examples of clean EEG signal, clean EMG signal, and noise-contaminated EEG signals

## 3.4 Training

### 3.4.1 Data processing

In this study, we synthesize EEG signals contaminated with EMG noise using EEG signal segments and EMG signal segments from the EEGdenoiseNet dataset. The clean EEG signal segments are linearly mixed with the EMG noise signal by Eq. (4) to generate the simulated noise-containing signal, see Fig. 2.

$$y = x + \lambda \cdot n \tag{4}$$

where y denotes the contaminated EEG signal, x denotes the clean EEG signal, n denotes the EMG noise signal and $\lambda$ is a parameter controlling the signal-to-noise ratio (SNR) level of the EMG noise to the EEG signal, and by adjusting $\lambda$, the SNR level of the noise-containing signal can be changed:

$$SNR = 10 \log \frac{RMS(x)}{RMS(\lambda \cdot n)} \tag{5}$$

The root mean square (RMS) value is defined as follows (Eq. (6)):

$$RMS(x) = \sqrt{\frac{1}{N} \sum_{i=1}^{N} a_i^2} \qquad (6)$$

where $a_i$ denotes the i-th signal segment, $N$ is the number of sampling points in the signal segment, and $N$ is 1024 for both EEG signal and EMG signal segments.

To make the number of EEG data consistent with the EMG data, we expanded the 4514 EEG signals to 5598 by randomly selecting the original EEG fragments for padding. The expanded EEG data is divided into 5000 training sets and 598 test sets, and the EMG signals are also divided, followed by the generation of contaminated signals. To ensure that the experimental results are realistic, the SNR range of the EMG noise-contaminated EEG signals is set to the usual range, -7 to 2 dB, and the contaminated signals are generated at each of the ten integer SNR levels to ensure the diversity and sufficiency of the dataset, resulting in a total of 55,980 contaminated signals at each SNR level. Subsequently, the noisy EEG time series signal fragments were converted into 32×32 pixel images for subsequent pix2pix model training. Fig. 3a and Fig. 3b show the clean EEG signal with the converted image, and Fig. 3c and Fig. 3d show the contaminated signal and its converted image.

After pix2pixGAN model processing, we get the generated denoised image and still need to convert the image back to time series. Firstly, the color image is converted to a grayscale image, and the minimum pixel value xmin and maximum pixel value xmax of the grayscale image $D$ are extracted for normalization. Subsequently, the pixel values of the grayscale image $D$ are normalized to the interval [0,1] by using Eq. (7) to obtain the normalized image $\tilde{D}$:

$$\tilde{D} = \frac{D - X_{\min}}{X_{\max} - X_{\min}} \qquad (7)$$

The normalized image $\tilde{D}$ is linearly mapped to the value domain of the original EEG signal to generate an image E that conforms to the EEG features, using Eq. (8):

$$E = \tilde{D} \times (y_{\max} - y_{\min}) + y_{\min} \qquad (8)$$

Through the above process, the image $\tilde{D}$ is scaled to the range $[y_{\min}, y_{\max}]$ of the EEG signal, and then the denoised EEG image E is reshaped into a one-dimensional time series. The generated denoised EEG signal is guaranteed to have the same amplitude characteristics as the actual EEG signal. Fig. 3e and Fig. 3f show the denoised signal and its corresponding image.

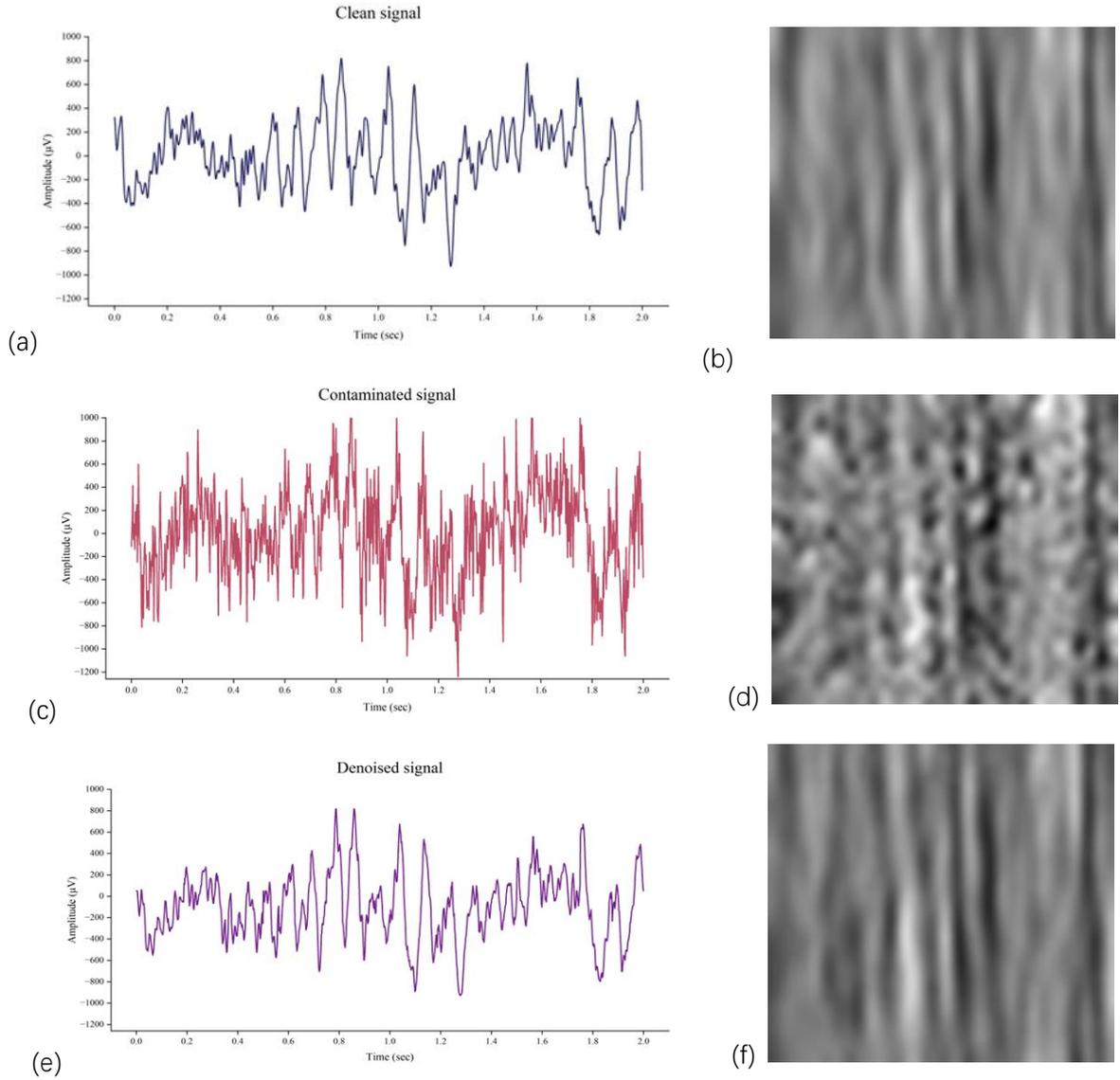

(a) (b) (c) (d) (e) (f)

Figure 3 Signals and corresponding converted images

### 3.4.2 Model training

The pix2pix GAN model used in this study is designed to realize the mapping transformation from noise-containing images to denoised images. In this experiment, the generator adopts the U-Net structure and its network architecture is shown in Fig. 4a. The size of the input image is 32×32 pixels, and the number of channels of the feature map is gradually increased from 64 to 256 after a series of convolutional layers and Batch Normalization layers are processed. The ReLU activation function is used in the generator, and six ResnetBlocks are introduced in the core part. Each residual block contains two convolutional layers and corresponding batch normalization layers and is implemented by Residual Connection, which effectively alleviates the gradient vanishing problem and facilitates the training of deep networks. Subsequently, the generator uses a transposed convolutional layer (ConvTranspose2d) for upsampling to recover the spatial dimension of the image gradually. Finally, after the Tanh activation function processing, the generator outputs a denoised signal image with a size of 32 ×

32 pixels.

The network structure of the discriminator is shown in Fig. 4b, which adopts the PatchGAN architecture to discriminate the local patches of the input image. The discriminator consists of four 1D convolutional layers, with the number of channels gradually increasing from 64 to 512, and a LeakyReLU activation function follows each convolutional layer. At the end of the network, a fully connected layer and a Sigmoid activation function are used to output the discriminative result, which is used to determine whether the input image patch is a real image or a generated image.

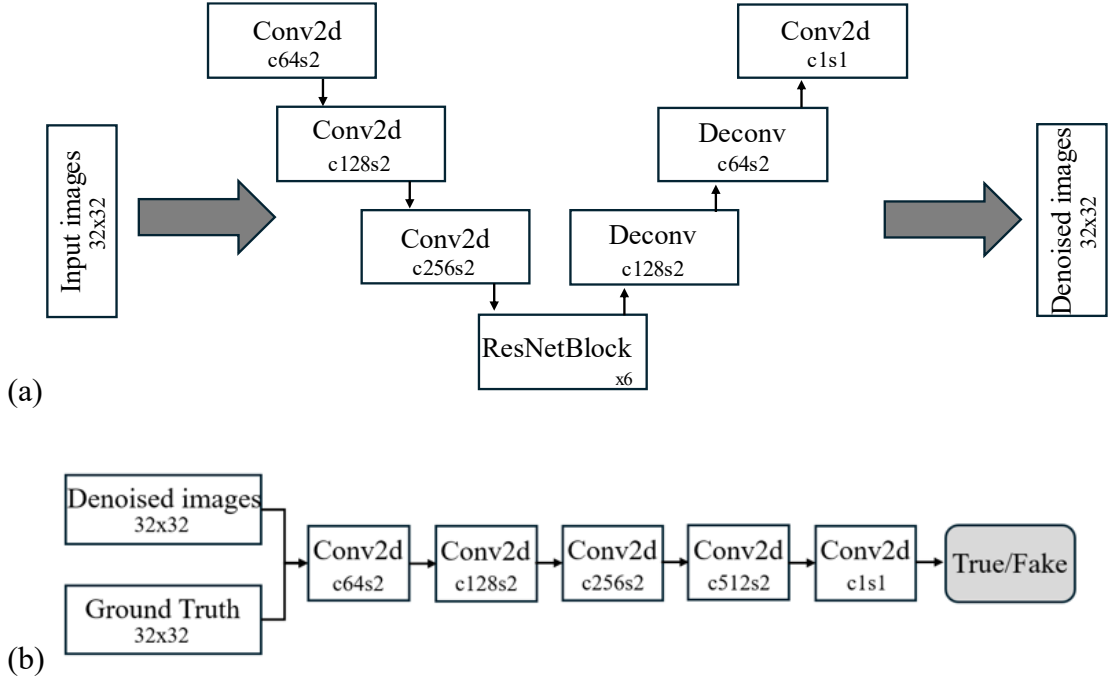

Figure 4 Generator's and Discriminator's network architectures

### 3.4.3 Performance Evaluation Metrics

In order to assess the denoising effect of the model in this study, we selected three key evaluation metrics: correlation coefficient (ACC), relative root mean square error in the time domain (RRMSE$_{temporal}$), and relative root mean square error in the frequency domain (RRMSE$_{spectral}$). In addition, we show the power spectral density of the signals to evaluate the differences between the signals from the frequency domain perspective, and thus to comprehensively assess the performance of the model.

The relative root mean square error in the time domain (RRMSE$_{temporal}$) is used to measure the error of the generated signal in the time domain and is defined in Eq. (9):

$$RRMSE_{temporal} = \frac{RMS\left(f\left(y\right) - x\right)}{RMS\left(x\right)} \qquad (9)$$

The frequency domain relative root mean square error (RRMSE$_{spectral}$) is used to evaluate the performance of the generated signal in the frequency domain, which is

defined in Eq. (10):

$$RRMSE_{spectral} = \frac{RMS\left(PSD\left(f\left(y\right)\right) - PSD\left(x\right)\right)}{RMS\left(PSD\left(x\right)\right)} \qquad (10)$$

where $f(y)$ denotes the generated denoised signal, $x$ is the original clean signal, and $PSD(\ )$ denotes the power spectral density of the signal. The correlation coefficient (ACC) is calculated by Eq. (11) and is used to measure the linear correlation between the generated signal and the original signal:

$$ACC = \frac{Cov\left(f\left(y\right), x\right)}{\sqrt{Var\left(f\left(y\right)\right)Var\left(x\right)}} \qquad (11)$$

Where the covariance $Cov\left(f\left(y\right), x\right)$ reflects the linear relationship between the generated signal $f(y)$ and the original clean signal $x$. $Var\left(f\left(y\right)\right)$ and $Var(x)$ denote the variance of the generated signal $f(y)$ and the real signal $x$, respectively.

# 4. RESULTS

In this section, we show the denoising performance of our deep learning model on the EEGdenoiseNet dataset. To give a general overview of the denoising results, Fig. 4 shows qualitative examples of our noise signal, original signal, and denoised signal in the time domain and after image transformation to better visualize our denoising model. Overall, in the removal of myogenic artifacts, the artifacts in the signals are cut down to a large extent and a noise-free EEG signal with good results is generated, especially at low signal-to-noise ratios, where a noise-free EEG signal with good results can still be obtained.

We quantitatively evaluate the strengths and weaknesses of the model performance employing three performance metrics (ACC, RRMSE$_{temporal}$, and RRMSE$_{spectral}$), and for a better evaluation, we use a total of ten integer SNR levels from -7 to 2. Fig. 5 illustrates the performance metrics evaluated for different deep learning models for EMG-contaminated EEG denoising, showing the ACC, RRMSE$_{temporal}$, and RRMSE$_{spectral}$ values obtained at ten integer SNR levels ranging from -7 to +2 dB. by comparing the performance of the model with other deep learning models, for all SNR levels, the Our pix2pixGAN network has excellent performance for EMG artifact removal, and for EMG artifact removal, as the SNR increases, the noise ratio decreases proportionally, the model's RRMSE$_{temporal}$ and RRMSE$_{spectral}$ values decrease, while the ACC increases, and the denoising performance rises. The pix2pixGAN model obtains substantially leading ACC values at lower signal-to-noise ratios, which indicates a higher correlation between the model-predicted EEG segments and the actual EEG segments, and at the same time, the pix2pixGAN model also has the lowest RRMSE values in both time and frequency domains, which further proves the superior performance of the model in removing EMG artifacts.

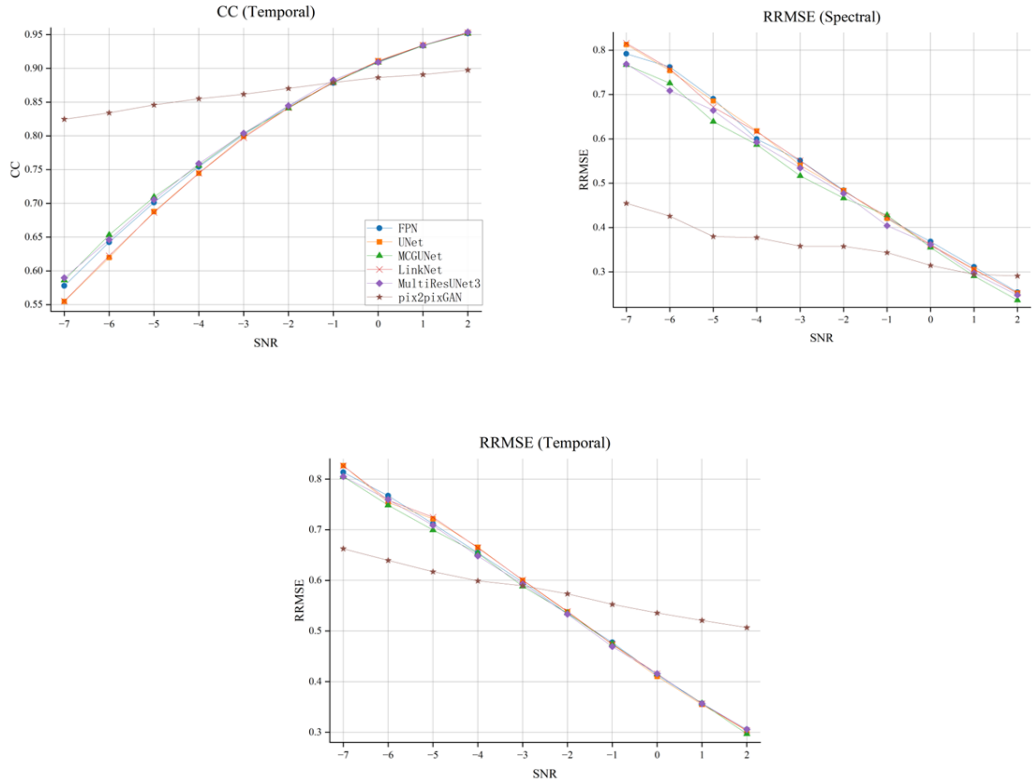

Figure 5 Metrics for Evaluating Different Models

We then further summarized the quantitative benchmarks by averaging all the performance metrics on signal-to-noise ratio for comparison. As can be seen in Table 1, the pix2pixGAN model has the lowest average $RRMSE_{spectral}$ values for removing myogenic artifacts and has the highest average CC value.

Table 1 Performance Comparison of Different Models

|  | CC | $RRMSE_{Temporal}$ | $RRMSE_{spectral}$ |
| --- | --- | --- | --- |
| FPN | 0.7995 | 0.5630 | 0.5234 |
| UNet | 0.7927 | 0.5647 | 0.5232 |
| MCGUNet | 0.8023 | **0.5573** | 0.5011 |
| LinkNet | 0.7924 | 0.5662 | 0.5239 |
| MultiResUNet3 | 0.8028 | 0.5594 | 0.5056 |
| pix2pixGAN | **0.8645** | 0.5797 | **0.3596** |

We also propose a performance metric to evaluate the denoising model, using the frequency power spectral density of the EEG to compare the performance of the model over different frequency bands. The entire frequency band ranges from 1 to 80 Hz, and the bands are subsequently divided into Delta (1-4 Hz), Theta (4-8 Hz), Alpha (8-13 Hz), Beta (13-30 Hz) and Gamma (30-80 Hz) bands. Fig. 6 and Table 2 show the corresponding PSD and power band ratios for specific EEG signals at -7 dB SNR, and it can be seen that with the addition of EMG noise, Beta's and gamma's power ratios in the frequency bands increase with the mixing of myogenic artifacts and decrease in the other bands, and in the denoised signals, the noise contamination from the myoelectric component in the frequency band has been removed. By showing the results at -7dB SNR, which is the theoretical worst case, it can be observed that our pix2pixGAN model can effectively remove the EMG noise from the EEG signal.

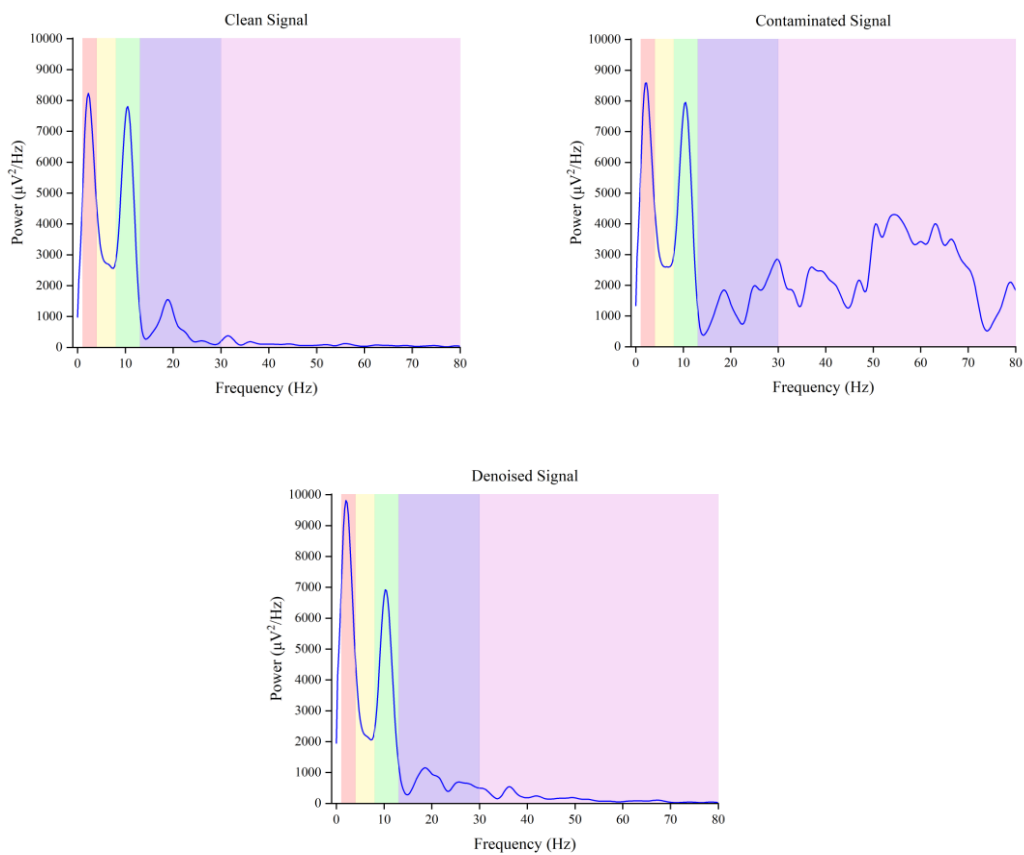

Figure 6 PSD of clean signal, contaminated signal, and denoised signal

Table 2 Power ratios of different frequency bands before and after noise removal.

|  | delta | theta | alpha | beta | Gamma |
|---|---|---|---|---|---|
| pix2pixGAN | 0.3314 | 0.1438 | 0.3024 | 0.1555 | 0.0952 |
| Ground truth | 0.2991 | 0.1767 | 0.3630 | 0.1294 | 0.0620 |
| Contaminated Signal | 0.1063 | 0.0592 | 0.1278 | 0.1200 | 0.6002 |

# 5. CONCLUSION

In our study, we introduce a novel deep learning model, the pix2pixGAN network, designed to effectively remove electromyographic artifacts (EMG noise) from electroencephalogram (EEG) data. EEG signals are usually subject to a variety of interferences, especially electromyographic artifacts, and these interferences significantly affect the quality of the signals and the subsequent analysis. Therefore, the development of an efficient denoising method has an important practical application value.

By using the EEG signals collected from the EEGdenoiseNet dataset with EMG signals, we generated noise-containing signals for model training. Experimental results show that our model performs well in denoising performance, with a correlation coefficient (ACC) as high as 0.8975 and a relative root mean square error ($RRMSE_{spectral}$) of 0.2909 in the frequency domain, which confirms its remarkable performance in the EEG denoising task. Further performance analysis and evaluation show that in a low signal-to-noise ratio (SNR) environment, my model still maintains a correlation of 0.8247 at -7 dB SNR, which significantly outperforms other deep-learning-based denoising models, proving its robustness in complex noise environments. These results validate the effectiveness of pix2pixGAN in processing EEG signals and also provide a novel and efficient solution for denoising EEG signals, which provides an important reference for future related research.